\documentclass[twocolumn,aps,showpacs]{revtex4}
\usepackage{epsfig}

\begin{document}
\voffset 1.8cm

\title{\LARGE \bf Excited bands in odd-mass rare-earth nuclei}

\author{Carlos E. Vargas}
\email{vargas@ganil.fr}
\affiliation{Grand Acc\'el\'erateur National d'Ions Lourds,
BP 5027, F-14076 Caen Cedex 5, France}
\author{Jorge G. Hirsch}
\email{hirsch@nuclecu.unam.mx}
\affiliation{Instituto de Ciencias Nucleares, Universidad
Nacional Aut\'onoma de M\'exico,Apartado Postal 70-543 M\'exico
04510 DF, M\'exico }
\author{Jerry P. Draayer}
\email{draayer@lsu.edu}
\affiliation{Department of Physics and Astronomy,
          Louisiana State University,
Baton Rouge, Louisiana 70803, U.S.A.}
\vskip 0.5cm
\date{\today}

\begin{abstract}
{\bf Abstract}
Normal parity bands are studied in $^{157}$Gd, $^{163}$Dy and
$^{169}$Tm using the pseudo SU(3) shell model. 
Energies and B(E2) transition strengths of states belonging to six  low-lying 
rotational bands with the same parity in each nuclei are presented.
The pseudo SU(3) basis includes states with pseudo-spin 0 and 1, and
${\frac 1 2}$ and ${\frac 3 2}$, 
for even and odd number of nucleons, respectively. States with pseudo-spin
1 and ${\frac 3 2}$ must be included for a proper description of some
excited bands.
\end{abstract}


\maketitle

\section{Introduction}

The nuclear shell model \cite{May49} is the cornerstone in the microscopic
study of nuclear properties. In the last two decades it has been
possible to perform calculations for valence nucleons in the full $sd-$ 
\cite{Bro88} and $fp-$ \cite{Cau95} shells, providing a detailed 
description of energy levels, and electromagnetic and weak transitions. 
Nevertheless, a shell model description of
heavy nuclei requires further assumptions that include a systematic and
proper truncation of the model space \cite{Val91}.

In light deformed nuclei the dominance of quadrupole-quadrupole interaction
led to the introduction of the SU(3) shell model \cite{Ell58}. However, the 
strong spin-orbit interaction renders the SU(3) truncation scheme useless in
heavier nuclei, while at the same time pseudo-spin emerges as a good symmetry 
\cite{Hec69}, and with it the pseudo-SU(3) model \cite{Dra82}. 

Pseudo-spin symmetry refers to the experimental fact that single-particle
orbitals with $j = l - {\frac 1 2}$ and $j = (l-2) + {\frac 1 2}$ in the 
shell $\eta$ lie very
close in energy, and can therefore be labeled as pseudo-spin doublets with
quantum   numbers $\tilde{\jmath} = j$, $\tilde\eta = \eta -1$, and $\tilde
l = l - 1$. The origin of this symmetry has been traced back to the
relativistic mean field equations \cite{Blo95}.

During the last decade the pseudo-SU(3) shell model has significantly evolved, 
becoming  a powerful microscopic theory for the description of the low energy 
rotational  bands in even-even heavy deformed nuclei, and of normal parity bands in 
heavy deformed odd-A nuclei. The first applications
considered pseudo-SU(3) a dynamical symmetry,
using only one irreducible representation (irrep) of SU(3) to describe the yrast
band up to the backbending regime \cite{Dra82}. After many improvements
in the rotor Hamiltonian, a technical breakthrough came with the
development of a computer code able to calculate reduced matrix elements of
physical operators between different SU(3) irreps \cite{Bah94}.
From then on it was possible to use a realistic Hamiltonian which, besides
the quadrupole-quadrupole interaction and rotor terms, includes the breaking
symmetry single particle energies and pairing interactions.

In this way a fully microscopic description of low-energy bands in even-even
and odd-A heavy deformed nuclei emerged. The model have been using as building
blocks the pseudo-SU(3) proton and neutron states having pseudo-spin 0 and
${\frac 1 2}$, which describe the even- and odd-A systems, respectively.  The
many-particle states are built as linear combinations of pseudo-SU(3) coupled
states with well-defined particle number and total angular momentum. Many
rotational bands and B(E2) electromagnetic transition strengths have been 
described in the even-even rare earth isotopes $^{156,158,160}$Gd,
$^{160,162,164}$Dy and $^{164,166,168}$Er \cite{Beu00,Pop00,Dra01} and in
the odd-mass $^{159,161}$Tb, $^{159,163}$Dy, $^{159}$Eu, $^{161}$Tm, and
$^{165,167}$Er nuclei \cite{Var00a,Hir00,Var00b,Var01}.

Widening the landscape of applications of the pseudo SU(3) model,
in the present work proton and neutron states having pseudo 
spin 1 and ${\frac 3 2}$, for even and odd number of particles, respectively,
are included in the Hilbert space.  The final many-particle states
in odd-mass nuclei have total spin ${\frac 1 2}$, ${\frac 3 2}$ or
${\frac 5 2}$. In this enriched space,
six or seven normal parity low-lying rotational bands in $^{157}$Gd, 
$^{163}$Dy and $^{169}$Tm are successfully described.
Many of them have important pseudo-spin 1 and ${\frac 3 2}$ components,
evidencing their relevance in the present study. Intra and inter-band
B(E2) transition strengths are also discussed.
This article complements the research of the scissors mode, M1 transition
strengths in the range between 2 and 4 MeV, performed in the same nuclei
\cite{Var02}.

In Section II a brief description of the pseudo SU(3) classification scheme is 
presented. The schematic Hamiltonian and its parametrization is
discussed in Section III. The results for low-lying rotational bands in
$^{157}$Gd, $^{163}$Dy and $^{169}$Tm are presented in Section IV, where
their wave functions are also discussed. The B(E2) intra- and inter-band 
transition strengths are presented in Section V. The summary and
conclusions are given in Section VI

\section{The Pseudo SU(3) Basis}

The starting point in any application of the pseudo SU(3) model is the 
building of the many-body basis. The proton and neutron valence Nilsson single 
particle levels are filled from below for a fixed deformation, allowing the 
determination of the occupancies in normal and unique parity orbitals \cite{Var00b}.
As it has been the case for all pseudo SU(3) studies up to now, nucleons in abnormal 
parity orbital are frozen, being the dynamics described using only nucleons in 
normal parity states. This choice is further reflected through the use of effective 
charges to describe quadrupole electromagnetic transitions which are larger than 
those usually employed in shell-model calculations.
While it has been shown that this is a reasonable and useful approach, it is 
nonetheless a strong assumption and the most important limitation of the present
model. A more sophisticated
treatment of the problem, with nucleons in intruder orbitals described in the
same footing using SU(3) irreps is under development \cite{Var98,Var01d}. 

In Table \ref{ocup} are presented the occupation numbers assigned to each nuclei.

\begin{table}[h]
\begin{center} \begin{tabular}{c|c|cccc}
Nuclei & $\beta$ & $n_\pi^N$ & $n_\nu^N$ & $n_\pi^A$ & $n_\nu^A$ \\ \hline
$^{157}$Gd & 0.250 & 8  & 7  & 6 & 4 \\
$^{163}$Dy & 0.258 & 10 & 9  & 6 & 6 \\
$^{169}$Tm & 0.267 & 11 & 12 & 8 & 6
\end{tabular}\end{center}
\caption{Deformation and occupation numbers.}
\label{ocup}\end{table}

The many-particle states of $n_\alpha$ active nucleons in a given normal
parity shell $\eta_\alpha$, $\alpha = \nu$ or $\pi$, can be classified by the
following chains of groups:
\begin{eqnarray}
~\{ 1^{n^{N}_\alpha} \} ~~~~~~~ \{ \tilde{f}_\alpha \} ~~~~~~~~\{ f_\alpha 
\} ~\gamma_\alpha ~ (\lambda_\alpha , \mu_\alpha ) ~~~ \tilde{S}_\alpha   
~~ \kappa_\alpha  \nonumber \\
~U(\Omega^N_\alpha ) \supset U(\Omega^N_\alpha / 2 ) \times U(2) \supset   
SU(3) \times SU(2) \supset \nonumber \\
\tilde{L}_\alpha  ~~~~~~~~~~~~~~~~~~~~~ J^N_\alpha ~~~~ \nonumber \\
SO(3) \times SU(2) \supset SU_J(2),
\label{chains}
\end{eqnarray}

\noindent where above each group the quantum numbers that characterize its 
irreps are given, and $\gamma_\alpha$ and $\kappa_\alpha$ are multiplicity
labels of the indicated reductions.

Any state $| J_i M \rangle$, where $J$ is the total angular momentum, $M$ its
projection and $i$ an integer index which enumerates the states with the same
$J, M$ starting from the one with the lowest energy, is built as a linear
combination 
\begin{equation}
| J_i M \rangle = \sum_\beta C^{J_i}_\beta |\beta J M \rangle \label{wf}
\end{equation}

\noindent of the strong coupled proton-neutron states
\begin{eqnarray}
|\beta JM \rangle \equiv \nonumber \\
| \{ \tilde{f}_\pi \} (\lambda_\pi \mu_\pi) \tilde{S}_\pi, \{ \tilde{f}_\nu \}
(\lambda_\nu \mu_\nu) \tilde{S}_\nu ; \rho (\lambda \mu ) \kappa L, \tilde{S}~ JM \rangle  =
\nonumber \\
  \sum_{M_L M_S} (L M_L, \tilde{S} M_S | J M ) 
\sum_{M_{S \pi} M_{S \nu}}
(\tilde{S}_\pi M_{S \pi}, \tilde{S}_\nu M_{S \nu} | \tilde{S} M_S)  \nonumber \\
\sum_{ \begin{array}{c} k_\pi L_\pi M_\pi \\ \kappa_\nu L_\nu M_\nu \end{array} }
{ \langle (\lambda_\pi \mu_\pi) \kappa_\pi L_\pi M_\pi ;
(\lambda_\nu \mu_\nu) \kappa_\nu L_\nu M_\nu |
(\lambda \mu ) \kappa L M \rangle}_\rho \nonumber\\
| \{ \tilde{f}_\pi \} (\lambda_\pi \mu_\pi) \kappa_\pi L_\pi M_\pi, \tilde{S}_\pi M_{S
\pi}  \rangle  ~~~~~~~~~~~~~~\label{basis}  \\
| \{ \tilde{f}_\nu \} (\lambda_\nu \mu_\nu) \kappa_\nu L_\nu M_\nu, \tilde{S}_\nu M_{S
\nu} \rangle  ~~~~~~~~~~~~~~~.\nonumber
\end{eqnarray}

In the above expression $\langle-;-|- \rangle$ and $(-,-|-)$ are the SU(3) and
SU(2) Clebsch Gordan coefficients, respectively.  

In all previous applications of the pseudo SU(3) model, only those
states with the highest spatial symmetry $\tilde{S}_{\pi,\nu}$ = 0 and ${\frac 1 2}$ 
have been included. In the present study states with 
$\tilde{S}_{\pi,\nu}$ = 1 and ${\frac 3 2}$ are also taken into account, allowing
for coupled proton-neutron states with total pseudo-spin
$\tilde{S}$ = ${\frac 1 2}$, ${\frac 3 2}$ or ${\frac 5 2}$. The pseudo-spin symmetry is close enough
to an exact symmetry in atomic nuclei, supporting a strong truncation
of the Hilbert space. However, pseudo spin-orbit partners are not exactly
degenerate, introducing a small degree of pseudo-spin mixing in the nuclear
wave function.

The quadrupole-quadrupole interaction can be expressed in terms of the
second order SU(3) Casimir operator $C_2$,
\begin{equation}
\hat{Q} \cdot \hat{Q} = 4 C_2 - 3 \hat{L}(\hat{L}+1).
\label{qq} \end{equation}

\noindent The  eigenvalue of $C_2$ for a given of SU(3) irrep
$(\lambda,\mu)$ is given by
\begin{equation}
\langle C_2 \rangle = {\lambda}^2 + {\mu}^2 + \lambda \mu + 3 \lambda + 
3 \mu  .
\end{equation}

\noindent The larger the expectation value of $C_2$, the greater the binding
of that SU(3) irrep by a pure attractive $Q\cdot Q$ interaction. The 
pseudo SU(3) basis is built
selecting those proton and neutron irreps with the largest $ \langle C_2
\rangle$ and pseudo-spin 0 and 1,  or ${\frac 1 2}$ and ${\frac 3 2}$, for even and odd number
of particles, respectively. The proton and neutron irreps are coupled to a total
pseudo SU(3) $(\lambda,\mu)$ irrep and to a total pseudo-spin $\tilde S = {\frac 1 2},
{\frac 3 2}$ and ${\frac 5 2}$. 
The basis states employed in the present work for each of the three  
nuclei studied are presented in Tables \ref{basis1}, \ref{basis2} and
\ref{basis3}.

\begin{table}[h]
\begin{center}
\begin{tabular}{ll|lllll}
 $(\lambda_\pi, \mu_\pi )S_\pi$ & $(\lambda_\nu, \mu_\nu )S_\nu$&
\multicolumn{3}{c}{total $(\lambda, \mu )$ } \\ \hline
  (10,4)0 & (15,5)${\frac 1 2}$ & (25,9)${\frac 1 2}$ & (27,5)${\frac 1 2}$ &(26,7)${\frac 1 2}$ \\
  (10,4)0 & (18,2)${\frac 1 2}$ & (27,5)${\frac 1 2}$ & (26,7)${\frac 1 2}$ &(29,4)${\frac 1 2}$ \\
          &   			& (28,6)${\frac 1 2}$ & (30,2)${\frac 1 2}$ \\
  (10,4)0 & (16,3)${\frac 1 2}$,${\frac 3 2}$ & (27,5)${\frac 1 2}$,${\frac 3 2}$ &(26,7)${\frac 1 2}$,${\frac 3 2}$ \\
  (10,4)0 & (17,1)${\frac 1 2}$,${\frac 3 2}$ & (27,5)${\frac 1 2}$,${\frac 3 2}$ \\
  (11,2)1 & (18,2)${\frac 1 2}$ & (27,5)${\frac 1 2}$,${\frac 3 2}$ & (29,4)${\frac 1 2}$,${\frac 3 2}$ \\
	  &			& (31,0)${\frac 1 2}$,${\frac 3 2}$ & (30,2)${\frac 1 2}$,${\frac 3 2}$ \\
  (11,2)1 & (15,5)${\frac 1 2}$ & (27,5)${\frac 1 2}$,${\frac 3 2}$ & (26,7)${\frac 1 2}$,${\frac 3 2}$ \\
  (11,2)1 & (16,3)${\frac 1 2}$,${\frac 3 2}$ & (27,5)${\frac 1 2}$,${\frac 3 2}$,${\frac 5 2}$ \\
  (12,0)0 & (15,5)${\frac 1 2}$ & (27,5)${\frac 1 2}$ \\
  (12,0)0 & (18,2)${\frac 1 2}$ & (30,2)${\frac 1 2}$ \\
  (8,5)0,1& (18,2)${\frac 1 2}$ & (26,7)${\frac 1 2}$,${\frac 3 2}$ & (27,5)${\frac 1 2}$,${\frac 3 2}$ \\
  (9,3)0,1& (18,2)${\frac 1 2}$ & (27,5)${\frac 1 2}$,${\frac 3 2}$
\end{tabular}\end{center}
\caption{The pseudo SU(3) irreps used in the description of
$^{157}$Gd.}\label{basis1}
\end{table}

\begin{table}[h]
\begin{center}
\begin{tabular}{ll|llll}
$(\lambda_\pi, \mu_\pi )S_\pi$ & $(\lambda_\nu, \mu_\nu )S_\nu$ &
\multicolumn{3}{c}{total $(\lambda, \mu )$ } \\ \hline
(10,4)0 & (16,7)${\frac 1 2}$ & (26,11)${\frac 1 2}$ \\
(7,7)0,1& (19,4)${\frac 1 2}$ & (26,11)${\frac 1 2}$,${\frac 3 2}$ \\
(10,4)0 & (19,4)${\frac 1 2}$ & (29,8)${\frac 1 2}$ & (31,4)${\frac 1 2}$ &
(30,6)${\frac 1 2}$ \\
	&		      & (32,2)${\frac 1 2}$ \\
(10,4)0 & (20,2)${\frac 1 2}$ & (31,4)${\frac 1 2}$ & (30,6)${\frac 1 2}$ &
(32,2)${\frac 1 2}$ \\
(10,4)0 & (21,0)${\frac 3 2}$ & (31,4)${\frac 3 2}$ \\
(11,2)1 & (19,4)${\frac 1 2}$ & (31,4)${\frac 1 2}$,${\frac 3 2}$ & (30,6)${\frac 1
2}$,${\frac 3 2}$ & (32,2)${\frac 1 2}$,${\frac 3 2}$ \\
(11,2)1 & (20,2)${\frac 1 2}$ & (31,4)${\frac 1 2}$,${\frac 3 2}$ & (32,2)${\frac 1
2}$,${\frac 3 2}$ \\
(12,0)0 & (19,4)${\frac 1 2}$ & (31,4)${\frac 1 2}$ \\
(11,2)1 & (21,0)${\frac 3 2}$ & (32,2)${\frac 1 2}$,${\frac 3 2}$,${\frac 5 2}$ \\
(12,0)0 & (20,2)${\frac 1 2}$ & (32,2)${\frac 1 2}$ 
\end{tabular}\end{center}
\caption{The pseudo SU(3) irreps used in the description of
$^{163}$Dy.}\label{basis2}
\end{table}

\begin{table}[h]
\begin{center}
\begin{tabular}{ll|lll}
$(\lambda_\pi, \mu_\pi )S_\pi$ & $(\lambda_\nu, \mu_\nu )S_\nu$ &
\multicolumn{3}{c}{total $(\lambda, \mu )$ } \\ \hline
(7,7)${\frac 1 2}$  & (24,0)0  & (31,7)${\frac 1 2}$ \\
(11,2)${\frac 1 2}$ & (20,5)0,1  & (31,7)${\frac 1 2}$,${\frac 3 2}$ & (33,3)${\frac 1
2}$,${\frac 3 2}$ & (32,5)${\frac 1 2}$,${\frac 3 2}$ \\
(7,7)${\frac 1 2}$  & (16,10)0 & (23,17)${\frac 1 2}$ \\
(11,2)${\frac 1 2}$ & (24,0)0  & (33,3)${\frac 1 2}$ & (34,1)${\frac 1 2}$ &
(35,2)${\frac 1 2}$ \\
(11,2)${\frac 1 2}$ & (21,3)0,1  & (33,3)${\frac 1 2}$,${\frac 3 2}$ & (32,5)${\frac 1
2}$,${\frac 3 2}$ & (34,1)${\frac 1 2}$,${\frac 3 2}$ \\
(11,2)${\frac 1 2}$ & (22,1)1  & (33,3)${\frac 1 2}$,${\frac 3 2}$ & (34,1)${\frac 1
2}$,${\frac 3 2}$ \\
(9,3)${\frac 1 2}$,${\frac 3 2}$  & (24,0)0  & (33,3)${\frac 1 2}$,${\frac 3 2}$ \\
(8,5)${\frac 1 2}$,${\frac 3 2}$  & (24,0)0  & (32,5)${\frac 1 2}$,${\frac 3 2}$ \\
(10,1)${\frac 1 2}$ & (24,0)0  & (34,1)${\frac 1 2}$ \\
(7,7)${\frac 1 2}$  & (20,5)0,1  & (27,12)${\frac 1 2}$,${\frac 3 2}$ \\
(11,2)${\frac 1 2}$ & (16,10)0 & (27,12)${\frac 1 2}$ \\
(4,10)${\frac 1 2}$ & (16,10)0 & (20,20)${\frac 1 2}$
\end{tabular}\end{center}
\caption{The pseudo SU(3) irreps used in the description of
$^{169}$Tm.}\label{basis3}
\end{table}

\section{The Pseudo SU(3) Hamiltonian}

The Hamiltonian contains spherical Nilsson single-particle terms for protons
($H_{sp,\pi}$) and neutrons ($H_{sp,\nu}$), the quadrupole-quadrupole
($\tilde Q \cdot \tilde Q$) and pairing interactions ($H_{pair,\pi}$ and
$H_{pair,\nu}$), as well as three `rotor-like' terms which are diagonal in the
SU(3) basis.

\begin{eqnarray}
 H & = & \sum_{\alpha =\pi, \nu} \{ H_{sp,\alpha} - ~ G_\alpha
~H_{pair,\alpha} \}
         - \frac{1}{2}~ \chi~ \tilde Q \cdot \tilde Q \label{ham} \\
   &   & + ~a~ K_J^2~ +~ b~ J^2~ +~ A_{sym}~
         \hat C_2 .\nonumber
\end{eqnarray}

This Hamiltonian can be separated into two parts: the first row includes
Nilsson single-particle energies and the pairing and quadrupole-quadrupole
interactions ($\tilde{Q}$ is the quadrupole operator in the pseudo SU(3)
space, see below). They are the basic components of any realistic
Hamiltonian and have been widely studied in the nuclear physics
literature, allowing their respective strengths to be fixed by systematics
\cite{Rin79,Duf96}. The SU(3) mixing is due to the single-particle and
pairing terms. The second row of the Hamiltonian (\ref{ham}) contains the
so-called
`rotor-like' terms, used to fine tune the moment of inertia and the
position of the different $K$ bands. They have been studied in detail in
previous papers where the pseudo SU(3) symmetry was used
as a dynamical symmetry \cite{Dra82}. The strength of these three terms were the only
ones adjusted nuclei by nuclei.
A detailed analysis of each term of this Hamiltonian and its parametrization 
has been presented elsewhere \cite{Var00b}.
 In Table \ref{para} are shown the
current values. These are the same used before for these three nuclei
\cite{tesis}, when the proton and neutron subspaces were reduced to S = 0 and
${\frac 1 2}$ (in other words, the same set of parameters are working in both
approximations of the theory).

\begin{table}
\begin{center} \begin{tabular}{c|cccccc}
Nuclei & a & b & A$_{sym}$ \\ \hline
$^{157}$Gd &  ~~ 0.046 & ~~ 0.0020 & ~~ 0.0008 \\
$^{163}$Dy & ~~ -0.040 & ~~ 0.0040 & ~~ 0.0016 \\
$^{169}$Tm &  ~~ 0.019 & ~~ 0.0007 & ~~ 0
\end{tabular}\end{center}
\caption{Parameters of the Hamiltonian (\ref{ham}).}
\label{para}
\end{table}

The electric quadrupole operator is expressed as \cite{Dra82}

\begin{equation}
Q_\mu = e_\pi Q_\pi + e_\nu Q_\nu \approx
e_\pi {\frac {\eta_\pi +1} {\eta_\pi}} \tilde Q_\pi +
e_\nu {\frac {\eta_\nu +1} {\eta_\nu}} \tilde Q_\nu , \label{q}
\end{equation}

\noindent
with effective charges $e_\pi = 2.3, ~e_\nu = 1.3$. These values
are very similar to those used in the pseudo SU(3) description  of
even-even nuclei \cite{Dra82,Beu98}. They are larger than those used in
standard calculations of B(E2) strengths \cite{Rin79} due to the passive
role assigned to the nucleons in unique parity orbitals, whose
contribution to the quadrupole moments is parametrized in this way.

The inclusion of configurations with pseudo-spin 1 and ${\frac 3 2}$ in
the Hilbert space  allows for a description of several highly excited
rotational bands in odd-mass nuclei. Their effect on the M1
transition strengths have been discussed in Ref. \cite{Var02}.
The most relevant result reported there is that, when the configuration
space is restricted to states with pseudo-spin 0
and ${\frac 1 2}$, it is not possible to find \cite{tesis} any M1
excitations with sizable strength from the ground state toward the states
between 2 to 4 MeV energy region for any of the three nuclei. The M1 
strength appears only when the Hilbert space is enlarged to include states
with $\tilde{S}_\pi$ or $\tilde{S}_\nu$ = 1 or 3/2.

\section{Rotational bands and spin content}

\begin{figure}[h]
\begin{center}
\vspace*{-1.8cm}
\leavevmode 
    \psfig{file=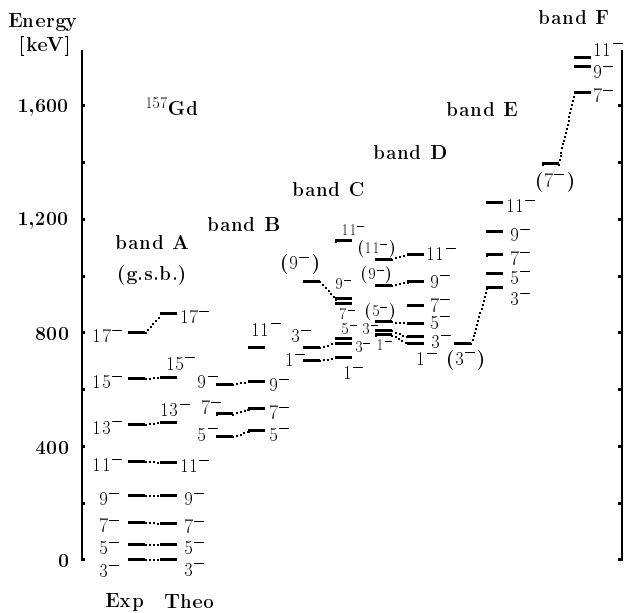,width=8.5cm}\vspace*{-2.2cm}
\caption{Negative parity bands in $^{157}$Gd. The integer numbers indicate twice 
the total angular momentum of each level.}
\label{sp-gd}
\end{center}
\end{figure}

Fig. \ref{sp-gd} shows the yrast and excited normal parity bands in
$^{157}$Gd. Experimental \cite{nndc} data are plotted on the left hand size
of each column, while those obtained using the Hilbert space and the
Hamiltonian parameters discussed in the previous sections are shown in the
right hand side. The agreement between both is excellent, but for the two
higher energy bands the lack of more experimental data in bands E and F (see
fig. \ref{sp-gd}) prevents a more rigorous comparison.

The whole energy spectra is built up by the interplay between the
single-particle and quadrupole-quadrupole terms in the Hamiltonian. The use of
realistic single-particle energies plays a key role in the appropriate
ordering of the different band-heads.

\begin{figure}[h]
\begin{center}
\vspace*{-1.8cm}
\psfig{file=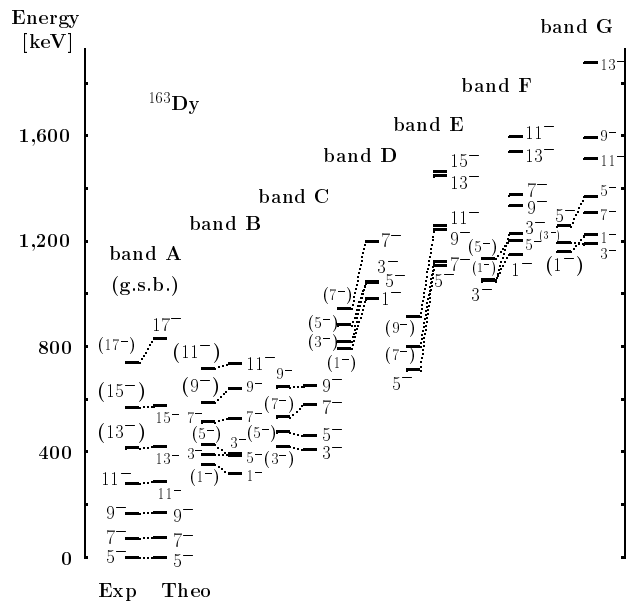,width=8.5cm}\vspace*{-2.2cm}
\caption{Negative parity bands in $^{163}$Dy.}
\label{sp-dy}
\end{center}
\end{figure}

\begin{figure}[h]
\vspace*{-1.8cm}
\psfig{file=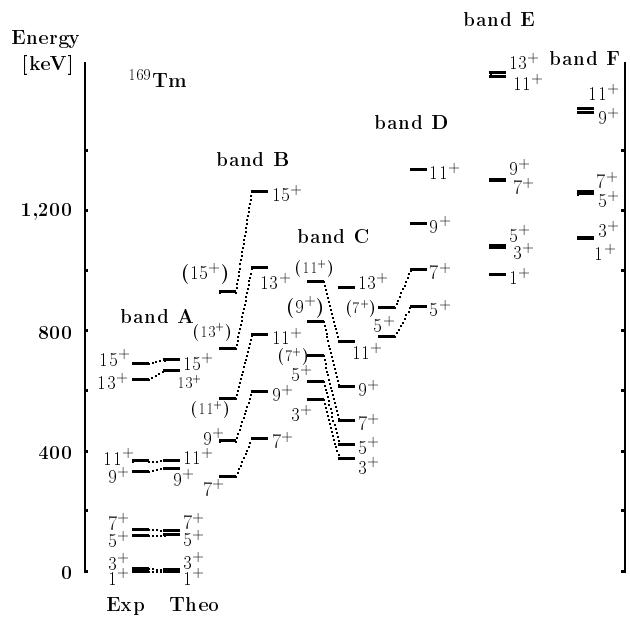,width=8.5cm}\vspace*{-2.2cm}
\caption{Positive parity bands in $^{169}$Tm.}
\label{sp-tm}
\end{figure}

Figs. \ref{sp-dy} and \ref{sp-tm} show the normal parity bands in $^{163}$Dy and
$^{169}$Tm, respectively. In $^{163}$Dy the agreement between theory and
experiment is very good for the seven rotational bands (A-G). These seven
bands represents nearly all the bands measured. For $^{169}$Tm the agreement between
the theoretical band structure and its experimental counterpart is still good,
although the differences in bandhead energies are larger than in the previous cases.

The bands plotted in Figs. (\ref{sp-gd}-\ref{sp-tm}) have a very regular
structure as one moves up the bands. The mixing remains nearly the same
for the states with
different angular moments belonging to the same band. As a
consequence, the pseudo-spin content is practically constant
along all the members of each band. Table \ref{spin-con} shows the
calculated pseudo-spin content of each rotational band in $^{157}$Gd,
$^{163}$Dy and $^{169}$Tm, respectively. As shown in Eq. (\ref{wf}), each
eigenstate is described as a linear combination of pseudo SU(3) states $
|\beta JM \rangle $, with pseudo-spin ${\frac 1 2}$, ${\frac 3 2}$ and
${\frac 5 2}$, depending on the proton and neutron pseudo-spin couplings.

\begin{table}
\begin{center} \begin{tabular}{cc|cc}
~~~Nuclei~~~ &~~~Band ~~~ & ~~~$\tilde{S}$ = 1/2 (\%)~~~ & ~~~$\tilde{S}$ = 3/2 (\%)~~~
\\ \hline
$^{157}$Gd& A & 89 & 11 \\
	  & B & 92 & 8 \\
	  & C & 92 & 8  \\
	  & D &  3 & 97 \\
       & * E & 77 & 23 \\
	  & F & 16 & 84 \\
	  &   &  &   \\
$^{163}$Dy& A & 100 &  0 \\
	  & B & 100 &  0 \\
	  & C & 100 & 0  \\
	  & D & 100 & 0  \\
	  & E & 0 & 100  \\
       & * F & 100 &  0 \\
	  & G & 0 & 100  \\
	  & & & \\ 
$^{169}$Tm& A & 93 & 7 \\
	  & B & 100  & 0  \\
	  & C & 9 & 91  \\
	  & D & 100 & 0  \\
	  & E & 35 & 65  \\
	  & F & 37 & 63  
\end{tabular}\end{center}
\caption{Pseudo-spin content for each band in $^{157}$Gd, $^{163}$Dy and
$^{169}$Tm. In first column it is indicated the nuclei, the second shows the
band, and third and four columns show the 1/2 and 3/2 pseudo spin-content 
as percentage for each band, respectively.}
\label{spin-con}
\end{table}

In most of the bands the mixing is very small, about half of them have no mixing
at all. The main contribution of the present work is to add those states with
pseudo-spin ${\frac 3 2}$ (and those marked with a *, see below) to the
pseudo SU(3) description of these nuclei. 

The total pseudo-spin content of the nuclear wave function is built with the coupling
of the $S_\pi$ and $S_\nu$ components. In all the cases, the states with 
pseudo-spin $\tilde S = {\frac 3 2}$ are built as $ 1 \otimes  {\frac 1 2}$, and
the $ 0 \otimes  {\frac 3 2}$ is not present in the bands analysed. 
By the other side, although most of the $\tilde S = {\frac 1 2}$ come from
the coupling $ 0 \otimes  {\frac 1 2}$, the bands marked with a * in Table \ref{spin-con},
are built as $\tilde{S}_\pi = 1  \otimes  \tilde{S}_\nu ={\frac 1 2}$ for the band  
E in $^{157}$Gd and $\tilde{S}_\pi ={\frac 1 2}  \otimes  \tilde{S}_\nu =1$ for 
the band F in $^{163}$Dy).

\begin{figure}[h]
\begin{center}
\vspace*{-2cm}
\hspace*{-0.6cm}
\leavevmode 
    \psfig{file=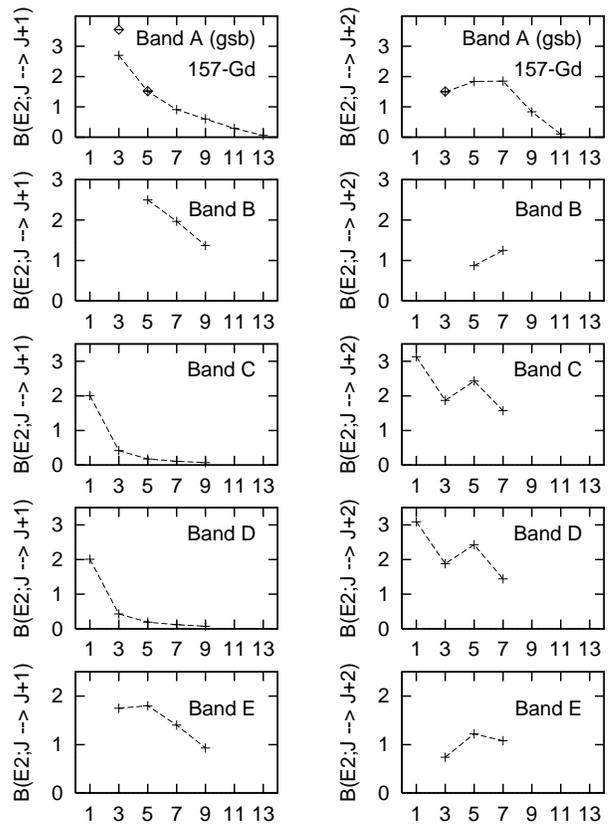,width=9.8cm}\vspace*{-1cm}
\caption{B(E2) intra-band transitions in $^{157}$Gd [$e^2 b^2 $].}
\label{be2-gd}
\end{center}
\end{figure}

\section{B(E2) transition strengths}

\begin{figure}[h]
\begin{center}
\vspace*{-1cm}
\hspace*{-1.6cm}
\leavevmode 
    \psfig{file=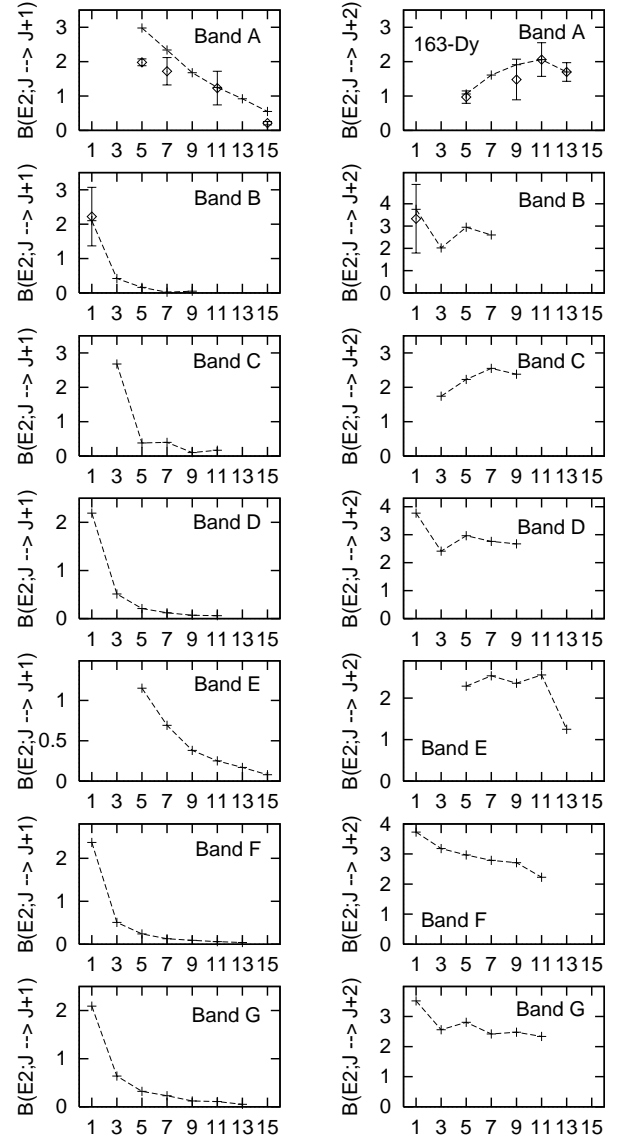,width=11.8cm}\vspace*{-0.1cm}
\caption{B(E2) intra-band transitions in $^{163}$Dy [$e^2 b^2 $].}
\label{be2-dy}
\end{center}
\end{figure}

\begin{figure}[h]
\begin{center}
\vspace*{-1cm}
\hspace*{-0.6cm}
\leavevmode 
    \psfig{file=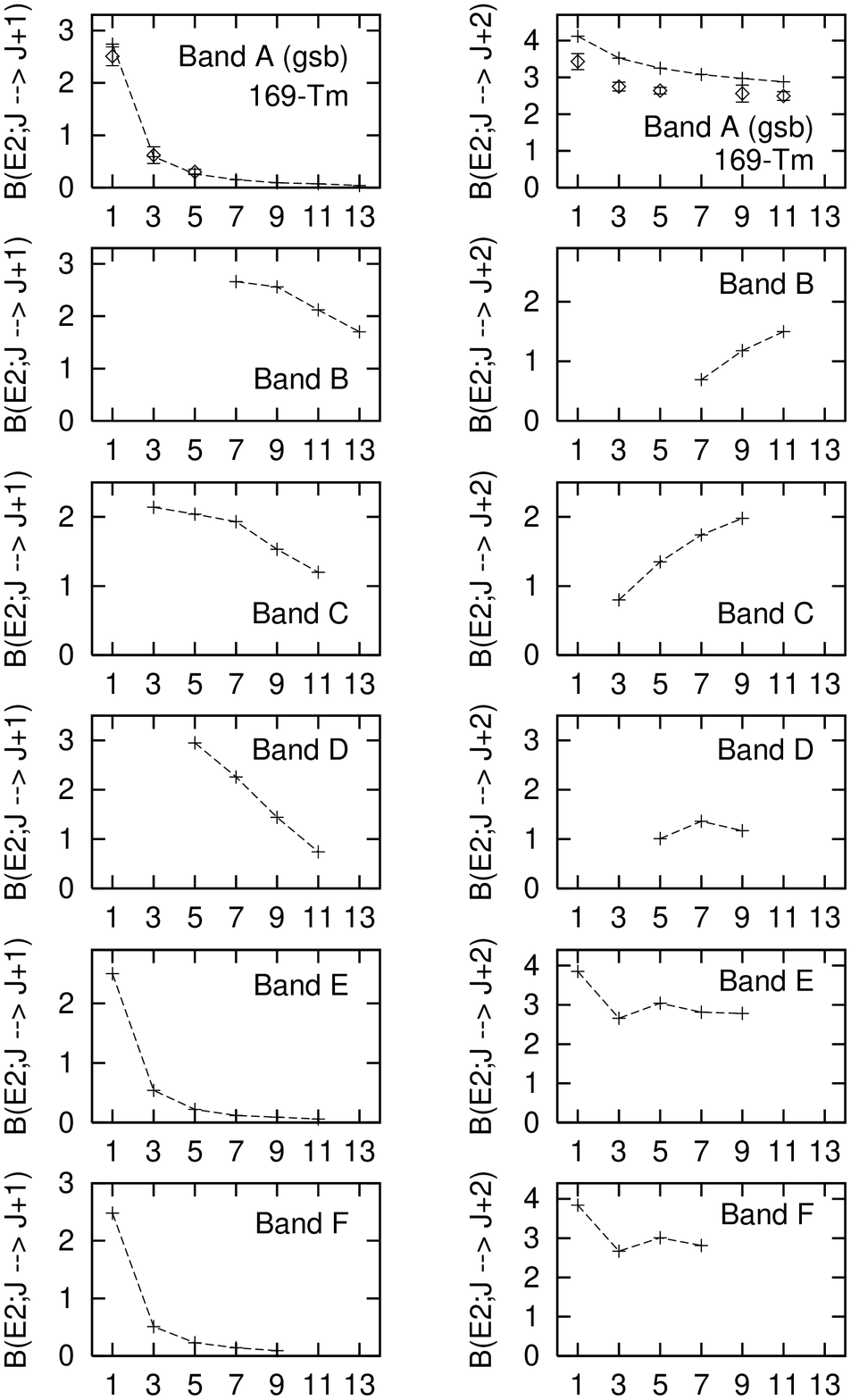,width=10cm}\vspace*{-0.1cm}
\caption{B(E2) intra-band transitions in $^{169}$Tm [$e^2 b^2 $].}
\label{be2-tm}
\end{center}
\end{figure}

In Figures \ref{be2-gd}, \ref{be2-dy} and \ref{be2-tm} are shown the B(E2)
intra-band transition strengths for the reported normal parity bands in
$^{157}$Gd, $^{163}$Dy and $^{169}$Tm, respectively. All experimental data 
available \cite{nndc} have been included with error bars for comparison.
For a given band, labeled A to G, the x axis represents twice the angular momentum
of the initial state for each transition, while the y axis shows 
the B(E2) strength. The graphs on the left hand side show B(E2;J $\rightarrow$ J+1)
transitions, and those in the right hand side present the B(E2;J $\rightarrow$ J+2),
both in units of $e^2 b^2$.
The ``+" symbols with segmented lines refers to the calculated values, 
and the diamonds with error bars to the experimental ones.

In Tables \ref{inter-gd}, \ref{inter-dy} and \ref{inter-tm} are shown many
inter-band B(E2) transition strengths in $^{157}$Gd, $^{163}$Dy and
$^{169}$Tm, respectively. All known experimental data available 
\cite{nndc} were included between parenthesis. Theoretical values are in
agreement with those reported previously for the lowest energy bands
in $^{163}$Dy \cite{Var01}.

The agreement between the calculated values with those measured is
remarkable. Some transitions, like 3/2$_C^-$ $\rightarrow$ 7/2$_D^-$,
3/2$_D^-$ $\rightarrow$ 7/2$_C^-$ in $^{157}$Gd, and
3/2$_B^-$ $\rightarrow$ 7/2$_C^-$, 3/2$_D^-$ $\rightarrow$ 7/2$_E^-$,
9/2$_C^-$ $\rightarrow$ 11/2$_B^-$ in  $^{163}$Dy, are  larger 
than 0.05 -- 0.10 $e^2 b^2$. It reflects the mixing between the different bands.

\begin{table}
\begin{center} \begin{tabular}{cccc}
$Ji_{band}^\pi \rightarrow Jf_{band}^\pi$ & B(E2)
& $Ji_{band}^\pi \rightarrow Jf_{band}^\pi$ & B(E2)\\ \hline
1/2$_C^-$ $\rightarrow$ 3/2$_A^-$ & 2.6 & 
1/2$_D^-$ $\rightarrow$ 3/2$_C^-$ & 8.5 \\
1/2$_C^-$ $\rightarrow$ 5/2$_B^-$ & 2.8 & 
1/2$_D^-$ $\rightarrow$ 5/2$_E^-$ & 2.1 \\
3/2$_C^-$ $\rightarrow$ 5/2$_D^-$ & 3.4 & 
3/2$_D^-$ $\rightarrow$ 5/2$_C^-$ & 3.1 \\
3/2$_C^-$ $\rightarrow$ $7/2_D^-$ &77.1 & 
3/2$_D^-$ $\rightarrow$ $7/2_C^-$ &77.8 \\
5/2$_D^-$ $\rightarrow$ $7/2_C^-$ & 2.1 & 
$7/2_A^-$ $\rightarrow$ $11/2_B^-$& 2.6 \\
$7/2_B^-$ $\rightarrow$ $11/2_A^-$& 2.1 & 
$7/2_D^-$ $\rightarrow$ $11/2_A^-$& 4.3 \\
$7/2_D^-$ $\rightarrow$ $11/2_B^-$& 6.6 & 
$7/2_D^-$ $\rightarrow$ $11/2_E^-$& 3.7 \\
$7/2_C^-$ $\rightarrow$ $11/2_D^-$&31.5 & 
$7/2_E^-$ $\rightarrow$ $11/2_D^-$&14.3 \\
$9/2_B^-$ $\rightarrow$ $11/2_A^-$& 3.1 & 
$9/2_E^-$ $\rightarrow$ $11/2_D^-$& 3.7 \\
$9/2_E^-$ $\rightarrow$ $11/2_C^-$& 3.3 &  
\end{tabular}\end{center}
\caption{Calculated B(E2) inter-band transitions in $^{157}$Gd [$e^2 b^2 \times
10^{-2}$].}
\label{inter-gd}
\end{table}

\begin{table}
\begin{center} \begin{tabular}{cccc}
$Ji_{band}^\pi \rightarrow Jf_{band}^\pi$ & B(E2) &
$Ji_{band}^\pi \rightarrow Jf_{band}^\pi$ & B(E2) \\ \hline
1/2$_B^-$ $\rightarrow$ 3/2$_C^-$ & 44.8 & 
1/2$_D^-$ $\rightarrow$ 3/2$_C^-$ &  2.6 \\
1/2$_F^-$ $\rightarrow$ 3/2$_G^-$ & 14.6 & 
1/2$_G^-$ $\rightarrow$ 3/2$_D^-$ &  5.6 \\
1/2$_G^-$ $\rightarrow$ 3/2$_F^-$ & 15.6 & 
1/2$_B^-$ $\rightarrow$ 5/2$_C^-$ &  3.4 \\ 
1/2$_B^-$ $\rightarrow$ 5/2$_A^-$ &  3.8 (4.0 $\pm$ 0.8) &
1/2$_D^-$ $\rightarrow$ 5/2$_A^-$ &  4.9 \\
1/2$_D^-$ $\rightarrow$ 5/2$_E^-$ &  1.2 & 
1/2$_F^-$ $\rightarrow$ 5/2$_E^-$ &  4.7 \\
1/2$_F^-$ $\rightarrow$ 5/2$_G^-$ &  1.6 & 
1/2$_G^-$ $\rightarrow$ 5/2$_E^-$ & 21.7 \\
3/2$_B^-$ $\rightarrow$ 5/2$_A^-$ &  2.3 (1.8 $\pm$ 0.6) & 
3/2$_C^-$ $\rightarrow$ 5/2$_B^-$ & 10.7 \\
3/2$_D^-$ $\rightarrow$ 5/2$_A^-$ &  1.6 & 
3/2$_D^-$ $\rightarrow$ 5/2$_C^-$ &  2.1 \\
3/2$_D^-$ $\rightarrow$ 5/2$_E^-$ & 32.1 & 
3/2$_G^-$ $\rightarrow$ 5/2$_E^-$ & 13.7 \\
3/2$_G^-$ $\rightarrow$ 5/2$_F^-$ &  2.8 & 
3/2$_F^-$ $\rightarrow$ 5/2$_G^-$ &  1.9 \\
3/2$_B^-$ $\rightarrow$ $7/2_A^-$ &  0.3 (3.7 $\pm$ 1.6) & 
3/2$_B^-$ $\rightarrow$ $7/2_C^-$ & 97.9 \\
3/2$_C^-$ $\rightarrow$ $7/2_A^-$ &  3.0 & 
3/2$_C^-$ $\rightarrow$ $7/2_B^-$ & 28.4 \\
3/2$_D^-$ $\rightarrow$ $7/2_A^-$ &  1.9 & 
3/2$_D^-$ $\rightarrow$ $7/2_B^-$ &  1.2 \\
3/2$_D^-$ $\rightarrow$ $7/2_E^-$ & 62.2 & 
3/2$_D^-$ $\rightarrow$ $7/2_G^-$ &  1.9 \\
3/2$_G^-$ $\rightarrow$ $7/2_E^-$ & 45.3 & 
3/2$_G^-$ $\rightarrow$ $7/2_F^-$ &  5.3 \\
3/2$_F^-$ $\rightarrow$ $7/2_G^-$ &  5.1 & 
5/2$_B^-$ $\rightarrow$ 5/2$_A^-$ &  0.4 \\
5/2$_B^-$ $\rightarrow$ $7/2_A^-$ &  2.0 (3.7 $\pm$ 2.1) & 
5/2$_B^-$ $\rightarrow$ $7/2_C^-$ & 11.5 \\
5/2$_E^-$ $\rightarrow$ $7/2_D^-$ &  6.3 & 
5/2$_E^-$ $\rightarrow$ $7/2_G^-$ & 45.4 \\
5/2$_B^-$ $\rightarrow$ $9/2_A^-$ &  2.2 (3.0 $\pm$ 1.6) &
5/2$_C^-$ $\rightarrow$ $9/2_A^-$ &  1.7  \\
5/2$_D^-$ $\rightarrow$ $9/2_A^-$ &  2.2  &
5/2$_G^-$ $\rightarrow$ $9/2_E^-$ &  3.2  \\
$7/2_C^-$ $\rightarrow$ $9/2_B^-$ & 11.4  &
$7/2_E^-$ $\rightarrow$ $9/2_D^-$ &  1.5  \\
$7/2_E^-$ $\rightarrow$ $9/2_F^-$ &  1.2  &
$7/2_C^-$ $\rightarrow$ $11/2_A^-$ & 2.2  \\
$7/2_E^-$ $\rightarrow$ $11/2_G^-$ &  2.0 &
$7/2_G^-$ $\rightarrow$ $11/2_E^-$ &  8.1 \\
$7/2_F^-$ $\rightarrow$ $11/2_G^-$ &  1.3 &
$9/2_B^-$ $\rightarrow$ $11/2_C^-$ &  4.7 \\
$9/2_C^-$ $\rightarrow$ $11/2_B^-$ & 63.8 &
$9/2_E^-$ $\rightarrow$ $11/2_G^-$ & 22.5 \\
$9/2_B^-$ $\rightarrow$ $13/2_A^-$ &  3.2 &
$9/2_C^-$ $\rightarrow$ $13/2_A^-$ &  1.3 \\
$9/2_D^-$ $\rightarrow$ $13/2_A^-$ &  1.9 &
$11/2_B^-$ $\rightarrow$ $13/2_C^-$ & 33.8 \\
$11/2_C^-$ $\rightarrow$ $13/2_A^-$ &  1.0 &
$11/2_C^-$ $\rightarrow$ $13/2_B^-$ &  7.5 \\
$11/2_C^-$ $\rightarrow$ $15/2_A^-$ &  1.7 &
\end{tabular}\end{center}
\caption{B(E2) inter-band transitions in $^{163}$Dy. The experimental data are
given between parenthesis[$e^2 b^2 \times
10^{-2}$].} \label{inter-dy}
\end{table}

\begin{table}
\begin{center} \begin{tabular}{cc}
 & \\
$Ji_{band}^\pi \rightarrow Jf_{band}^\pi$ & B(E2) \\ \hline
1/2$_E^+$ $\rightarrow$ 3/2$_C^+$ &  1.2 \\
1/2$_E^+$ $\rightarrow$ 3/2$_F^+$ &  7.9 \\
1/2$_F^+$ $\rightarrow$ 3/2$_C^+$ &  1.1 \\
1/2$_F^+$ $\rightarrow$ 3/2$_E^+$ &  9.0 \\
1/2$_E^+$ $\rightarrow$ 5/2$_C^+$ &  1.0 \\
3/2$_A^+$ $\rightarrow$ 5/2$_C^+$ &  0.004 \\
3/2$_C^+$ $\rightarrow$ 5/2$_A^+$ &  0.107 \\
3/2$_E^+$ $\rightarrow$ 5/2$_C^+$ &  0.011 \\
3/2$_E^+$ $\rightarrow$ 5/2$_D^+$ &  0.001 \\
3/2$_E^+$ $\rightarrow$ 5/2$_F^+$ &  3.5 \\
3/2$_F^+$ $\rightarrow$ 5/2$_C^+$ &  1.5 \\
3/2$_F^+$ $\rightarrow$ 5/2$_E^+$ &  1.7 \\
3/2$_A^+$ $\rightarrow$ $7/2_B^+$ &  0.00168 (0.0006 $\pm$ 0.0001) \\
3/2$_A^+$ $\rightarrow$ $7/2_C^+$ &  0.012 \\
3/2$_A^+$ $\rightarrow$ $7/2_D^+$ &  0.029 \\
3/2$_C^+$ $\rightarrow$ $7/2_A^+$ &  0.015 \\
3/2$_C^+$ $\rightarrow$ $7/2_D^+$ &  0.005 \\
3/2$_F^+$ $\rightarrow$ $7/2_C^+$ &  0.9 \\
5/2$_D^+$ $\rightarrow$ $7/2_B^+$ &  1.1 \\
5/2$_A^+$ $\rightarrow$ $7/2_B^+$ &  0.0037 (0.0014 $\pm$ 0.0002) \\ 
5/2$_E^+$ $\rightarrow$ $7/2_C^+$ &  1.1 \\
5/2$_E^+$ $\rightarrow$ $7/2_F^+$ &  2.4 \\ 
5/2$_F^+$ $\rightarrow$ $7/2_E^+$ & 23.6 \\
5/2$_D^+$ $\rightarrow$ $9/2_B^+$ &  0.9 \\ 
5/2$_E^+$ $\rightarrow$ $9/2_C^+$ &  1.1 \\
$7/2_A^+$ $\rightarrow$ $7/2_B^+$ &  0.00611 (0.00176 $\pm$ 0.00005) 
\end{tabular}\end{center}
\caption{B(E2) inter-band transitions in $^{169}$Tm. Experimental data
available are between parenthesis [$e^2 b^2 \times
10^{-2}$]}.\label{inter-tm}
\end{table}

The B(M1;$J_{g.s.}^\pi \rightarrow J_f^\pi$) transitions in $^{157}$Gd,
$^{163}$Dy and $^{169}$Tm have been presented in Ref. \cite{Var02}. 
$J_{g.s.}^\pi$ refers to the ground states ${\frac 3 2}^-$, ${\frac 5 2}^-$ and 
${\frac 1 2}^+$, respectively, in these nuclei. In that work it was shown
that most of the states with
energies between 2 and 4 MeV have very important contributions from
states with proton and neutron pseudo-spin 1 and ${\frac 3 2}$.

\section{Summary and conclusions}

The pseudo SU(3) shell model for odd-mass nuclei has been shown to
offer a quantitative microscopic description of several normal parity
rotational bands in $^{157}$Gd, $^{163}$Dy and $^{169}$Tm. The present
article complements the study of the scissors mode in these nuclei,
their fragmentation and their summed B(M1;$\uparrow$) strengths.

In order to successfully describe the excited bands, intra- and inter-band
B(E2) and B(M1) transition strengths, it was necessary to enlarge the
Hilbert space, including those pseudo
SU(3) irreps with the largest $C_2$ values having pseudo-spin
1 and ${\frac 3 2}$ and to use realistic values for the single-particle
energies. This expansion of the model space allowed
for the description of new low lying normal parity bands, in 
agreement with the experimental data. The new bands have predominantly 
$\tilde{S} = {\frac 3 2}$, but it was shown that the most important
contribution comes from the proton or neutron subspaces with $\tilde{S} =
1$. It implies that the pseudo-spin mixing in the wave function takes
place mostly in the sub-space with even number of particles.
The interplay between the single-particle and the quadrupole-quadrupole
terms in the Hamiltonian defines this mixing. The delicated balance
between these two interactions, defines the gross features of the
calculated excited bands.

Intra- and inter-band B(E2) transition strengths are predicted in good
agreement with their measured values and with those reported previously
with the pseudo SU(3) model. Detailed description of magnetic properties
of these rotational bands \cite{Stu02} and the excited bands in
other rare-earth nuclei are desirable.

\section{Acknowledgements}
This work was supported in part by CONACyT (M\'exico) and the US National
Science Foundation.


\begin{thebibliography}{99}

\bibitem{May49} M. G. Mayer, Phys. Rev. {\bf 75}, 1969 (1949); O. Haxel,
	J. H. D. Jenssen, and H. E. Suess, Phys. Rev. {\bf 75}, 1766 (1949).
\bibitem{Bro88} B. A. Brown and B .H. Wildenthal, Ann. Rev. Nucl. Part. Sc.
	{\bf 38}, 29 (1988).
\bibitem{Cau95} E. Caurier, {\it et. al.} Phys. Rev. Lett. {\bf 75}, 2466
	(1995); A. Schmidt, {\it et. al.} Phys. Rev. C {\bf 62}, 044319 (2000).
\bibitem{Val91} M. Valli\'eres and H. Wu, in {\it Computational Nuclear
	Physics 1} edited by K. Langanke, J. A. Maruhn, and S. E. Koonin
	(Springer, Berlin, 1991).
\bibitem{Ell58} J. P. Elliott, Proc. Roy. Soc. London Ser. A {\bf 245},
	128 (1958); {\bf 245}, 562 (1958).
\bibitem{Hec69} K. T. Hecht and A. Adler, Nucl. Phys. {\bf A 137}, 129
	(1969); A. Arima, M. Harvey, and K. Shimizu, Phys. Lett. {\bf B
	30}, 517 (1969).
\bibitem{Dra82} J. P. Draayer, {\it et. al.}, Nucl. Phys. A {\bf 381}, 1
	(1982); J. P. Draayer and K. J. Weeks, Ann. of Phys. {\bf 156}, 41 
	(1984); O. Casta\~nos, {\it et. al.}, Ann. of Phys. {\bf 329},
	290 (1987).
\bibitem{Blo95} A. L. Blokhin, {\it et. al.}, Phys. Rev.
	Lett. {\bf 74}, 4149 (1995); J. N. Ginocchio, Phys. Rev. Lett. {\bf
	78}, 436 (1997); J. Meng, {\it et. al.} Phys. Rev. {\bf C 58},
	R632 (1998).
\bibitem{Bah94} C. Bahri and J.P. Draayer, Comput. Phys. Commun. {\bf
	83}, 59 (1994).
\bibitem{Beu00} T. Beuschel, J.G. Hirsch, and J.P. Draayer,
	{ Phys. Rev.} {\bf C 61}, 54307 (2000).
\bibitem{Pop00} G. Popa, J. G. Hirsch and J. P. Draayer,
	{ Phys. Rev.} {\bf C 62}, 064313 (2000).
\bibitem{Dra01} J. P. Draayer, G. Popa and J. G. Hirsch,
	{Acta Phys. Pol.} {\bf B 32}, 2697 (2001).
\bibitem{Var00a} C. Vargas, J. G. Hirsch, T. Beuschel, J. P. Draayer,
	{ Phys. Rev. } {\bf C 61}, 31301 (2000).
\bibitem{Hir00} J. G. Hirsch, C.E. Vargas, and J. P. Draayer,
	{Rev. Mex. Fis.} {\bf 46} Supl. 1, 54 (2000).
\bibitem{Var00b} C.E. Vargas, J. G. Hirsch and J.P. Draayer,
	{Nucl. Phys.} {\bf A 673}, 219-237 (2000).
\bibitem{Var01} C. Vargas, J. G. Hirsch, J. P. Draayer,
	{Phys. Rev.} {\bf C 64}, 034306 (2001).
\bibitem{Var02} C. E. Vargas, J. G. Hirsch, J. P. Draayer,
	Subbmited to Phys. Rev. Lett.
\bibitem{Var01d} C. E. Vargas, J. G. Hirsch, J. P. Draayer, Nucl. Phys. {\bf
	A 690}, 409 (2001); {\it ibid} , Nucl. Phys. {\bf A 697}, 655 (2002)
\bibitem{Var98} C. Vargas, J. G. Hirsch, P. O. Hess, and J. P. Draayer, Phys.
	Rev. {\bf C 58}, 1488 (1998);
	J. P. Draayer, and K. J. Weeks, Ann. Phys. {\bf 156}, 41 (1984).
\bibitem{Rin79} P. Ring and P. Schuck. {\it The Nuclear Many-Body
	Problem}, Springer, Berlin  (1979).   
\bibitem{Duf96} M. Dufour and A. P. Zuker, Phys. Rev. {\bf C 54}, 1641 
	(1996).
\bibitem{tesis} PhD Thesis, Carlos E. Vargas, CINVESTAV 2001, M\'exico.
\bibitem{Beu98} T. Beuschel, J.P. Draayer, D. Rompf , and J.G. Hirsch,
	Phys. Rev. {\bf C 57}, 1233 (1998).
\bibitem{nndc} National Nuclear Data Center,
	http:\verb+//+bnlnd2.dne.bnl.gov
\bibitem{Stu02} A. Stuchbery, Nucl. Phys. {\bf A 700}, 83 (2002).

\end{thebibliography}
\end{document}